\documentstyle[12pt]{article}

\makeatletter
\def\case{\protect\@case}

\def\@case#1#2{%
\def\@tempa{#2}\def\@tempb{/}%
\ifx\@tempa\@tempb %
\def\@tempa{\@@case{#1}}%
\else %
\def\@tempa{\@@case{#1}{#2}}%
\fi
\@tempa
}
\def\@@case#1#2{{\textstyle{#1\over#2}}}

\def\text#1{%
\relax
\ifmmode %
\mathchoice
{\hbox{\everymath{\displaystyle}\rm #1}}%
{\hbox{\everymath{\textstyle}\rm #1}}%
{\hbox{\everymath{\scriptstyle}%
\def\prm{\fam\z@ \the\scriptfont\z@ \relax}%
\def\pit{\fam\itfam \the\scriptfont\itfam \relax}%
\rm #1}%
}%
{\hbox{\everymath{\scriptscriptstyle}%
\def\prm{\fam\z@ \the\scriptscriptfont\z@ \relax}%
\def\pit{\fam\itfam \the\scriptscriptfont\itfam \relax}%
\rm #1}%
}%
\else %
\leavevmode\hbox{#1}%
\fi
}

\makeatother

\begin{document}
\pagestyle{myheadings}

\title{\bf Physical Consequences of the Interpretation of the Skew
Part of {\boldmath${g_{\mu\nu}}$} in Einstein's Nonsymmetric Unified
Field Theory\thanks{{\em Australian Journal of
Physics}, {\bf 48}, No.~1, 45--53, (1995).}}

\author{Joseph Voros \\
{\small\sl Department of Physics, Monash University,
Clayton, Victoria, 3168, Australia.}}

\date{}

\maketitle
\thispagestyle{empty}

\begin{abstract}
The electromagnetic interaction in the Einstein-Infeld-Hoffmann (EIH)
equations of motion for charged particles in Einstein's Unified Field
Theory is found to be {\em automatically\/} precluded by the
conventional identification of the skew part of the fundamental tensor
with the Faraday tensor.  It is shown that an alternative
identification, suggested by observations of Einstein, Bergmann and
Papapetrou, would lead to the expected electromagnetic interaction,
were it not for the intervention of an infelicitous (radiation) gauge.
Therefore, an EIH analysis of EUFT is {\em inconclusive\/} as a test
of the physical viability of the theory, and it follows that EUFT
cannot be considered necessarily unphysical on the basis of such an
analysis.  Thus, historically, Einstein's Unified Field Theory was
rejected for the wrong reason.
\end{abstract}

\section{Introduction}

Recently there has been considerable discussion on the nonsymmetric
gravitational theory (NGT) of Moffat (Moffat 1979, 1991).  Damour {\em
et~al} (1992, 1993) have assailed NGT, claiming it to be theoretically
inconsistent and possessing unphysical behaviour, and Moffat and his
collaborators have defended NGT  (Moffat 1993; Cornish and Moffat
1993, 1994; Cornish {\em et~al\/} 1993).  This debate is still open
and, owing to the formal mathematical similarity of NGT to the
nonsymmetric unified field theory of Einstein (1950, 1956), this
discussion has some bearing on Einstein's theory, and thereby
rekindles interest in what many have considered a closed subject.
While NGT presumes the skew part of the nonsymmetric fundamental
tensor to be a new, unknown field with possible novel couplings to
matter, Einstein's theory assumes it to be of electromagnetic origin.

Einstein's unified field theory (Einstein 1950, 1956) was considered
untenable owing to its apparent failure to produce correct equations
of motion for charged particles.  This apparent untenability
stimulated consideration of various modifications to the theory  (Bose
1953; Bonnor 1954; Moffat and Boal 1975; Klotz 1982; Antoci 1989),
including the recent work of Damour {\em et~al\/} (1992,1993) and that
on NGT (Moffat 1979, 1991, 1993; Cornish and Moffat 1993, 1994;
Cornish {\em et~al\/} 1993).  A comprehensive review and general
outline of the early work was given by Goenner (1984).

The ostensible problem in EUFT is that charged particles do not appear
to feel the electromagnetic field, a conclusion reached by Infeld
(1950) for the earlier, and Callaway (1953) for the later version.
Each investigator used a modified form of the Einstein-Infeld-Hoffmann
(EIH) approximation scheme (Einstein and Infeld 1949) which was
developed to find the equations of motion of masses in General
Relativity (GR) from the free-space field equations alone.  It turns
out that the two versions of EUFT are essentially equivalent insofar
as equations of motion are concerned, so we may confine our attention
to the earlier, upon which Infeld's analysis is based.

The purpose of this paper is to show that an EIH analysis of EUFT is
{\em inconclusive} as a test of the physical viability of the theory.
There is {\em no} desire to demonstrate that the theory {\em is
viable} --- the purpose is merely to show that an EIH analysis is
unable to conclude one way or the other.

In the case of the Infeld-Callaway analysis, the conventional
interpretation of the $f_{\mu\nu}$ as the Faraday tensor causes the
Coulomb interaction to vanish automatically due to ``extra''
derivatives inherent in the interpretation.  This would seem to
suggest that the $f_{\mu\nu}$ would need to be interpreted as
potentials rather than as Faraday tensor type derivatives of
potentials, if there is to be any prospect of a Coulomb interaction in
the EIH equations of motion for EUFT.

In fact, such an interpretation of the $f_{\mu\nu}$ is suggested by
three separate observations: (i) by the precise way that the EIH
equations of motion arise from the field equations of GR,
Einstein-Maxwell Theory (EMT) and EUFT; (ii) by the intimation of a
gauge-fixing role for Eq.~(\ref{Gamf}) below by Einstein
(1956) and Bergmann (1956) and (iii) by an analysis due
to Papapetrou (1948).  Under this interpretation of $f_{\mu\nu}$
as potentials, EUFT in its original form would yield the Coulomb
interaction in the EIH equations of motion, were it not for the fact
that EUFT is put into the radiation gauge by the condition
(\ref{Gamf}).

Since the EIH scheme produces a similarly vanishing Coulomb
interaction when applied to EMT in the radiation gauge, the vanishing
of the Coulomb term in this gauge for EUFT is not sufficient to
conclude that EUFT is therefore necessarily unphysical, any more than
the same situation in EMT is sufficient to conclude that EMT is
unphysical --- one recognizes the lack of a Coulomb interaction to
result from the choice of {\em gauge}.  Of course, this is very
different from asserting that EUFT {\em is\/} physical, which is not
the intention.  The purpose here is merely to show that, since an EIH
analysis is inconclusive as a test of the viability of EUFT,
historically, EUFT, rejected as it was on the basis of such an
analysis, was rejected for the wrong reason.

\section{Field Equations}

In EUFT, both the fundamental tensor $g_{\mu\nu}$ and the connection
$\Gamma^\alpha_{\mu\nu}$ are assumed to be non-symmetric.  The field
equations are (Einstein 1950; Infeld 1950)
\begin{eqnarray}
   g_{\mu\nu,\lambda} &=&  g_{\sigma\nu}\Gamma^\sigma_{\mu\lambda}
        +g_{\mu\sigma}\Gamma^\sigma_{\lambda\nu} , \label{gGam} \\
\Gamma_\mu &\equiv& \case{1}{2}
(\Gamma^\tau_{\mu\tau} - \Gamma^\tau_{\tau\mu}) = 0,
  \label{Gammamu} \\
R_{(\mu\nu)} &=& 0,    \label{R50sy} \\
R_{[\mu\nu]} &=& 0;    \label{R50sk} \\
\text{where}\quad R_{\mu\nu} &=&
  \partial_\sigma\Gamma^\sigma_{\mu\nu}
- \case{1}{2}\bigl(\partial_\nu\Gamma^\sigma_{(\mu\sigma)}
+ \partial_\mu\Gamma^\sigma_{(\nu\sigma)}\bigr) \nonumber\\
 &\quad& - \Gamma^\sigma_{\mu\tau}\Gamma^\tau_{\sigma\nu}
+ \Gamma^\sigma_{\mu\nu}\Gamma^\tau_{(\sigma\tau)} .
 \label{R50eqn}
\end{eqnarray}
In this paper, Greek letters denote spacetime indices (0,1,2,3) and
Latin letters denote spatial indices (1,2,3).  Round (square)
brackets around indices denote symmetry (skew-symmetry) as usual.

The symmetric part of the nonsymmetric fundamental tensor is
identified with the metric tensor of GR, and the skew part
$f_{\mu\nu}$ was conventionally identified with the Faraday field
tensor of classical electromagnetism.   Eq.~(\ref{Gammamu}) gives rise
to the equivalent equation
\begin{equation}
\left(\sqrt{-g}f_{\mu\sigma}\right)^{,\sigma}=0,
\label{Gamf}
\end{equation}
where $g=\det(g_{\mu\nu})$, which thus has the form of Maxwell equations.

It was the failure of EUFT, with this interpretation of $f_{\mu\nu}$,
to produce the Coulomb interaction in the EIH equations of motion
which led to the widely-held view (Misner {\em et~al\/}
1973) that EUFT is unphysical and which prompted numerous
modifications of the theory  (Goenner 1984).

\section{EIH Equations of motion}

The equations of motion for EUFT are obtainable using the
Einstein-Infeld-Hoffmann (EIH) approximation scheme as employed by
Infeld (1950) (and Callaway (1953)).  Here we simply give an outline
of the basic method.  Full details may be found elsewhere (Einstein
and Infeld 1949; Wallace 1940, 1941; Scheidegger 1953).

The EIH scheme was originally developed to answer the question of
whether the free-space field equations of GR were sufficient to
produce the equations of motion of particles which were not test
particles, i.e.\@ which were also a source of the field (the geodesic
principle applies to test particles only).  Einstein suspected the
non-linearity of the field equations of GR might give rise to
constraints on the motions of particles, and thereby yield the
equations of motion.

The field equations are split, with the time component distinguished.
The basic assumption is of ``slow'' motion with respect to the speed
of light, which may be easily formalised, in terms of an expansion
parameter based on the ``speed of motion'' rather than on ``strength
of field.''  Particles are represented as singularities in the field.
Since the field equations do not therefore hold at the positions of
the particles, they are each surrounded by a closed surface upon which
the field equations do hold (and each surface encloses only one
particle).  The field equations are integrated, and it turns out that
the values of the surface integrals are independent of the shapes of
the surfaces --- they depend only on the co-ordinates of the
singularities and their time derivatives.  One finds that in order for
the whole system of equations to remain consistent at each successive
instant of time, the surface integrals must take certain values.
Since the value of the surface integrals depends only upon the motion
of the enclosed singularities, the singularities are thereby
effectively constrained to move in certain ways.  In other words, the
surface integrals imply integrability conditions and these conditions
are the equations of motion of the particles.

The Newtonian equations of motion arise in the first iteration of the
EIH approximation, at the fourth order in the expansion parameter.
When there is an electromagnetic energy-momentum tensor in the field
equations, the correct charge-particle equations of motion emerge.

\subsection{EIH for GR and EMT}

In the EIH method, the character of the entire approximation scheme
depends upon the choice of solution of the lowest-order equations.  No
use is made of the more familiar exact solutions of the field
equations, such as the Schwarzschild in GR or Reissner-Nordstr\"om in
EMT --- they have no role whatsoever in the EIH scheme.  The EIH
approximation scheme is independent of the results of the (full
non-linear) field equations, such as these exact solutions.  The point
of contact between the field equations and the EIH scheme lies in the
assumed character of the field functions.

Thus in GR, where the gravitational functions (i.e. the symmetric
$g_{\mu\nu}$) are interpreted as potentials, the lowest-order (i.e.
second order) EIH functions were chosen to be Newtonian potentials
$\varphi$, with $\varphi$ being the total sum of each particle's own
$m/r$-type Newtonian gravitational potential.

In EMT, the electromagnetic functions $A_\mu$ are interpreted as
electromagnetic potentials.  To the lowest order (second), one
obtains simply the Coulomb potential $\Phi$, with $\Phi$ being the
total sum of each particle's own $q/r$-type Coulomb potential.

In the ``Newtonian'' approximation, the EIH equations of motion for
EMT arise (Wallace 1941) from a surface integral whose integrand is
formed from the fourth-order part of
\begin{equation}
P_{ij} +\case{1}{2}\delta_{ij}\eta^{\alpha\beta}P_{\alpha\beta} + 2
(T_{ij} +\case{1}{2}\delta_{ij}\eta^{\alpha\beta}T_{\alpha\beta}) = 0,
\label{modGR}
\end{equation}
where $P_{\mu\nu}$ is the usual Ricci tensor of GR, $T_{\mu\nu}$ is
the usual electromagnetic energy-mo\-men\-tum tensor, and
$\eta_{\mu\nu}$ is the flat space metric, $\text{diag}(1,-1,-1,-1)$.
The integrand contains derivatives and/or products of the second order
Newtonian and Coulomb potentials.  The GR equations of motion are
obtained by putting $T_{\mu\nu}=0$.

For later comparison, we show the result for the simplest case --- two
particles --- and give only the equations of motion for particle ``1.''
Those of particle ``2'' are analogous.  The equations of motion of
particle ``1'' are found to be (in 3-vector form)
\begin{equation}
\stackrel{1}{m}\ddot{\mbox{\boldmath$x$}} + 2\stackrel{1}{m}
\nabla\left[\stackrel{2}{m}/r\right] = 0
 \label{newteqm}
\end{equation}
where $r$ is the distance between the two masses, \mbox{\boldmath$x$}
denotes the position 3-vector of particle ``1,'' dots denotes time
derivatives, and $\stackrel{1}{m}$ and $\stackrel{2}{m}$ denotes the
masses of particles ``1'' and ``2'' respectively.  We see that this is
the Newtonian equation of motion for particle ``1'' moving in the
gravitational field of particle ``2.''.

The modification to the gravitational equations of motion
(\ref{newteqm}), brought about by the electromagnetic field in
(\ref{modGR}), is found from the surface integral of the
non-gravitational terms. These are (Wallace 1940, 1941)
\begin{equation}
- 4 \Phi_{,i}\Phi_{,j} + 2 \delta_{ij}\Phi_{,s}\Phi_{,s}\;,
\label{wallaceintegrand}
\end{equation}
and yield a term of the form
\begin{equation}
\stackrel{1}{q} \nabla\left[\stackrel{2}{q}/r\right]
\label{coulterm}
\end{equation}
which is clearly the Coulomb force on particle ``1'' (with charge
$\stackrel{1}{q}$) due to the field of particle ``2'' (with charge
$\stackrel{2}{q}$).

\subsection{EIH for EUFT}

In EUFT, the modification to the fourth-order gravitational equations
of motion due to the electromagnetic field is found in a way exactly
analogous to EMT.  The $f_{\mu\nu}$ were conventionally interpreted as
Faraday tensor derivatives of the electromagnetic 4-potential.  The
EIH integrand is formed from the fourth-order part of (\ref{R50sy}) by
way of
\begin{equation}
R_{(ij)} +\case{1}{2}\delta_{ij}\eta^{\alpha\beta}R_{(\alpha\beta)}
= 0,
\end{equation}
and contains the gravitational part of (\ref{modGR}) (the $P$'s) which
yield the same equations of motion (\ref{newteqm}) as in GR, together
with non-gravitational terms containing products of the $f_{\mu\nu}$
and its derivatives.

The EUFT modification to the gravitational equations of motion is
found by evaluating the surface integral of these non-gravitational
terms, as in EMT, although in EUFT, these terms appear in an integrand
more complicated than that for EMT.

This was the basis of Infeld's approach.  However, Infeld did not
fully simplify his non-gravitational integrand and, by use of the
so-called Lemma of the EIH procedure (a formal algebraic result), he
was able to correctly conclude, {\em without evaluating any of the
terms}, that the integral vanished.  This use of the Lemma, however,
obscures what turns out to be an important fact.

Simplifying Infeld's integrand (Infeld 1950) we find it is
\begin{equation}
\Phi_{,is}\Phi_{,js} - \Phi_{,s}\Phi_{,ijs},
 \label{infeldintegrand}
\end{equation}
which is to be compared with (\ref{wallaceintegrand}).  The
gravitational equations of motion in EUFT are thereby modified
(Voros 1994) by a term of the form
\begin{equation}
\stackrel{1}{q} \nabla\left[\nabla^2(\stackrel{2}{q}/r)\right]
\label{uftcoul}
\end{equation}
The ``extra'' derivatives in (\ref{infeldintegrand}) compared to
(\ref{wallaceintegrand}) result in the appearance of the Laplacian
operator $\nabla^2$ in the EUFT analogue (\ref{uftcoul}) of the EMT
Coulomb force term (\ref{coulterm}) of the equations of motion.  Since
$q/r$ is a harmonic function, this term will vanish identically {\em
even if the Coulomb potential does not\/} and there is therefore no
electromagnetic modification to the gravitational equations of motion,
whence Infeld's (and thus Callaway's) inference that charged particles
do not feel the electromagnetic field.  Infeld's use of the Lemma
meant he did not observe this.  However, with the details of the
calculation laid bare in this way we see that any possible Coulomb
contribution to the EIH equations of motion is {\em automatically}
precluded, entirely on account of these ``extra'' derivatives, which
are engendered by the conventional identification of $f_{\mu\nu}$ with
{\em derivatives\/} of electromagnetic potentials.

These ``extra'' derivatives would be avoided by an identification of
the $f_{\mu\nu}$ with potentials, which is consonant with the
following observations of Einstein, Bergmann and Papapetrou.

Einstein (1956) noted that the curvature tensor, formed as
usual from the $\Gamma^\alpha_{\mu\nu}$, is invariant under the
substitution (``$\lambda$-transformation'')
\begin{equation}
\Gamma^\alpha_{\mu\nu} \rightarrow \Gamma^\alpha_{\mu\nu} +
 \delta^\alpha_\mu \lambda_{,\nu}
\end{equation}
where $\lambda$ is an arbitrary function of the coordinates.  This
$U(1)$ invariance he termed ``$\lambda$-invariance.''   The
non-symmetric Ricci tensor formed as usual from the curvature tensor
is also invariant under a $\lambda$-transformation.  Einstein then
noted that postulating the equations (\ref{Gammamu}) involves a
normalization of the $\Gamma$-field, which removes the
$\lambda$-invariance of the system of equations --- the non-symmetric
Ricci tensor reduces to the $R_{\mu\nu}$ defined in Eq.~(\ref{R50eqn})
if Eq.~(\ref{Gammamu}) is assumed.

Bergmann (1956) noted that $\lambda$-transformations are related
to electromagnetic $U(1)$ gauge transformations.  Thus (\ref{Gammamu})
implies by way of (\ref{Gamf}), that the $f_{\mu\nu}$ are potentials,
since gauge conditions are generally of the nature of
single-derivative constraints.

The possibility of interpreting the $f_{\mu\nu}$ as potentials was
explicitly noted earlier by Papapetrou (1948), who analysed the
field equations of EUFT in order to compare them with those of EMT.
He concluded that the simplest interpretation of $f_{\mu\nu}$ was not
as the field itself, but as the potential of the field.  The apparent
concern, that the skew tensor $f_{\mu\nu}$ is of a different character
than the vector potential $A_\mu$ of Maxwell's electromagnetic theory,
and thus that it might not be able to give rise to the correct number
of degrees of freedom for a photon field, when examined in detail,
turns out to be unfounded (see Comments).

As we noted earlier, in GR, where the gravitational functions are
interpreted as potentials, the lowest-order EIH functions were chosen
to be Newtonian potentials.  In EMT, the electromagnetic functions are
interpreted as electromagnetic potentials, and this gives rise, at the
lowest order, to the Coulomb potential.  Thus, also, in EUFT, informed
by the interpretation of $f_{\mu\nu}$ as potentials, the lowest order
electromagnetic functions in the $f_{\mu\nu}$ must correspond to
Coulomb potentials.  This must be done in a way which follows the
identification made for the metric in GR and EMT i.e. that the lowest
order $f_{\mu\nu}$ be treated as a set of Coulomb potentials in
accordance with the manner in which the lowest-order functions in the
metric are treated as a set of Newtonian potentials in the EIH
approach to GR and EMT.

When this is done, we find (Voros 1994) that in EUFT the
gravitational equations of motion are modified by a term of the form
\begin{equation}
\stackrel{1}{q} \nabla(\stackrel{2}{q}/r) .
\label{coul}
\end{equation}
The term above is to be compared with Eq.~(\ref{coulterm}).  It is
clearly the Coulomb force on particle $1$ moving in the field of the
other particle.

It follows from (\ref{coul}), therefore, that a Coulomb term {\em
may\/} exist in the EIH equations of motion of EUFT in the
interpretation of $f_{\mu\nu}$ as potentials, in contrast to the
situation under the conventional interpretation as Faraday tensor type
derivatives of potentials.

However, Equation~(\ref{Gamf}), which has been identified, following
the observations of Einstein and Bergmann, as a gauge-fixing equation,
can be shown to take the form $\Phi=0, \nabla \cdot
\mbox{\boldmath$A$} = 0$ (to third order in the EIH scheme) which
specifies the radiation gauge.

In EMT, imposing the radiation gauge on the EIH scheme causes all the
particle charges to be constrained to vanish.  Since all the
electromagnetic EIH functions depend on the charges, it follows that
if the radiation gauge is imposed, then all the EIH functions vanish
to all orders, the scheme ``collapses,'' and is thus inapplicable even
in the case of EMT. There is nothing mysterious about this --- the EIH
scheme was developed to describe the interaction of {\em particles};
it is ill-suited to describing {\em radiation}.

It therefore follows that this gauge constraint in EUFT, similarly
constraining all charges to be zero, precludes a charged particle
interaction in the EIH equations of motion for EUFT.  There is thus no
Coulomb interaction in the equations of motion, despite the provision
for one in (\ref{coul}), because it is proscribed by the gauge
entailed by Eq.~(\ref{Gamf}).

\section{Comments}

Some comments on the interpretation of $f_{\mu\nu}$ as potentials
outside the context of the EIH approximation scheme are in order.

At first glance it may seem unlikely that the skew tensor $f_{\mu\nu}$
could contain the degrees of freedom of the two helicity states of a
massless spin-1 particle; one might be tempted to think that because
the theory has a 2-form playing the role of gauge potentials, it might
describe instead an axion field or something similar.  However, such a
conclusion is premature.  The actual behaviour of the $f_{\mu\nu}$, at
least in the linearized case, is very suggestive, as we now report.

A spin-projection analysis of the implied particle spectrum performed
recently by Moffat (1993) on the linearized skew field equations of
NGT (which has similar field equations to the later Einstein theory,
and identical linearized skew free-space field equations to those of
the later EUFT), reveals that the six functional degrees of freedom in
the skew $f_{\mu\nu}$ actually behave like two spin-1 fields (i.e.
each having three functional degrees of freedom), both massless.  It
turns out that one of these does not propagate, so the six
$f_{\mu\nu}$ effectively possess the same behaviour as a single
propagating massless spin-1 field.

However, this spin-1 field is further constrained to have only a
scalar degree of freedom, due to the presence of the constraint
equation (\ref{Gamf}) in the theory.  [It can be shown that in the
classical linearized skew field equations, the apparently six degrees
of freedom in the $f_{\mu\nu}$ are reduced, by the linearized form of
the constraint (\ref{Gamf}), to a single degree of freedom.]

Moffat interprets this one remaining scalar degree of freedom as
indicating the skew field to be a scalar field;  however, we can see
from the details of Moffat's result, that the linearized skew
free-space field equations of EUFT for the $f_{\mu\nu}$ may also be
interpreted as a constrained propagating massless spin-1 field.
Indeed, by adding mass to the skew field $f_{\mu\nu}$ of their theory,
Damour {\em et~al\/} (1993) are able to explicitly extract all three
spin-1 degrees of freedom, whence their interpretation of $f_{\mu\nu}$
as a massive fifth-force-type vector field.

What role the constraint equation (\ref{Gamf}) has in EUFT remains an
interesting question --- within the limited context of the EIH
approximation, it behaves like a gauge-fixing equation.  It would have
a more complex role outside this approximation.

\section{Conclusion}

The equation (\ref{Gamf}) reduces to Maxwell equations given the {\em
a priori\/} identification of $f_{\mu\nu}$ with the electromagnetic
(Faraday) field tensor.  We have seen that this identification
frustrates the inference of a Coulomb interaction via the EIH
procedure owing to ``extra'' derivatives on the Coulomb potential
introduced by this identification.  The (EIH independent) gauge-fixing
character of Eq.~(\ref{Gamf}) to which Einstein and Bergmann allude,
however, suggests an alternative identification of $f_{\mu\nu}$ with
electromagnetic potentials, which was independently noted earlier by
Papapetrou to be the simplest interpretation of the $f_{\mu\nu}$.

Given this identification of $f_{\mu\nu}$ as potentials,
Eq.~(\ref{Gamf}) imposes, within the context of the EIH scheme, a
gauge which precludes the existence of a charged particle interaction
in the equations of motion --- the radiation gauge.   The absence of a
Coulomb interaction in EMT in the radiation gauge is not usually taken
to mean that EMT is therefore physically unviable.  It is clear that
the absence of such an interaction in this circumstance for Einstein's
theory also does not allow the standard inference that EUFT is
therefore necessarily unphysical owing to the lack of such an
interaction, which latter was the contention of Infeld (1950) and
Callaway (1953).

Thus, an EIH analysis of EUFT is inconclusive as a test of the
physical viability of the theory.  This is not to say, of course,
that the theory is viable --- we have merely found that an EIH
analysis which, historically, was the basis of rejecting EUFT, is
unable to determine the physical viability of EUFT one way or the
other.

Given that the skew field $f_{\mu\nu}$ has been found to behave like a
constrained massless propagating spin-1 field in the linearized
approximation, one is led to wonder what might emerge from Einstein's
Unified Field Theory if the constraint equation (\ref{Gamf}) on the
$f_{\mu\nu}$ were to be relaxed.

\section*{Acknowledgements}

The author would like to thank Dr.~H.S.~Perlman for his inestimable
support, for a critical reading of the manuscript and for many helpful
discussions and suggestions, and to the Referee for pertinent
observations and helpful suggestions. This work was supported in part
by an Australian Postgraduate Research Award.


\section*{References}

\strut\indent Antoci, S.  (1989), {\em Gen. Rel. Grav.} {\bf 21},~171--83.

Bergmann, P.  (1956), {\em Phys. Rev.} {\bf 103},~780--1.

Bonnor, W.  (1954), {\em Proc. Roy. Soc. (London) A}
  {\bf 226},~366--76.

Bose, S.  (1953), {\em J. Phys. Rad.} {\bf 14},~641--4.

Callaway, J.  (1953), {\em Phys. Rev.} {\bf 92},~1567--70.

Cornish, N. and Moffat, J. (1993), {\em Phys. Rev. D} {\bf 47},~4421--4.

Cornish, N. and Moffat, J.  (1994),
  {\em Nonsymmetric gravity does have acceptable global asymptotics.}
  Preprint gr-qc/9401018.

Cornish, N., Moffat, J. and Tatarski, D.  (1993), {\em Phys. Lett. A}
{\bf 173},~109--115.

Damour, T., Deser, S. and McCarthy, J.  (1992), {\em Phys. Rev D}
{\bf 45},~R3289--91.

Damour, T., Deser, S. and McCarthy, J.  (1993), {\em Phys. Rev D}
{\bf 47},~1541--56.

Einstein, A.  (1950), {\em The Meaning of
  Relativity}, third edn, Princeton University Press, Princeton, New Jersey,
  USA.

Einstein, A.  (1956), {\em The Meaning of
  Relativity}, fifth edn, Princeton University Press, Princeton, New Jersey,
  USA.

Einstein, A. and Infeld, L.  (1949), {\em Can. J. Math.} {\bf 1},~209--41.

Goenner, H.  (1984), Unified field theories:
 {} From Eddington and Einstein up to now, {\em in} V.~D. Sabbata and
  T.~Karade, eds, `Sir Arthur Eddington Centenary Symposium, Vol. 1 --
  Relativistic Astrophysics and Cosmology', World Scientific, Singapore,
  pp.~176--96.

Infeld, L.  (1950), {\em Acta Phys. Polonica} {\bf 10},~284--93.

Klotz, A.  (1982), {\em Macrophysics and
  Geometry}, Cambridge University Press, Cambridge, UK.

Misner, C.W., Thorne, K.S., and Wheeler, J.A. (1973),
  {\em Gravitation}, W.H. Freeman \& Company, San Francisco, USA.

Moffat, J.  (1979), {\em Phys. Rev. D} {\bf 19},~3554--8.

Moffat, J.  (1991), Review of the nonsymmetric
  gravitational theory, {\em in} R.~Mann and P.~Wesson, eds, `{\em
  Gravitation 1990}, Proceedings of the Banff Summer Institute on Gravitation,
  Banff, Alberta, August 12--25, 1990', World Scientific, Singapore, p.~523.

Moffat, J.  (1993), {\em Consistency of the
  nonsymmetric gravitational theory.}
University of Toronto Preprint UTPT-93-11.

Moffat, J. and Boal, D.  (1975),
  {\em Phys. Rev. D} {\bf 11},~1375--82.

Papapetrou, A.  (1948), {\em Proc. Roy. Irish
  Acad.} {\bf 52A},~69--86.

Scheidegger, A.E.  (1953) {\em Rev. Mod. Phys.} {\bf 25},~451--68.

Voros, J. (1994),
Proceedings of the Fourth Monash General Relativity Workshop, July
1993, Department of Mathematics, Monash University (eds A.~Lun,
L.~Brewin and E.~Chow), p122.  Published by Monash University, Clayton,
Victoria, 3168, Australia. ISBN:~0~7326~0554~7.

Wallace, P.  (1940), PhD thesis, Department of
  Applied Mathematics, University of Toronto, Canada.

Wallace, P.  (1941), {\em Am. J. Math.} {\bf 63},~729--40.

\end{document}